\documentclass{aaushoot}
\usepackage{graphicx,natbib}
\newcommand{\nablaad}{\nabla_{\mbox{\scriptsize ad}}}

\begin{document}
\title{A new model for the lower overshoot layer in the Sun}
\author{D. Marik$^{1,2}$\and K. Petrovay$^1$}
\authorrunning{Marik \and Petrovay}

\offprints{D. Marik}
\institute{$^1$E\"otv\"os University, Department~of Astronomy, Budapest, Pf.~32, 
              H-1518 Hungary\\
$^2$SPARC, Dept. of Applied Mathematics, Univ. Sheffield, Hicks
Bldg., Hounsfield Rd., S3 7RH Sheffield, UK.\\
              \email{d.marik@astro.elte.hu}
             }

   \date{}

\abstract{
We present a model for the lower overshoot layer of  the Sun, based on the
realistic solar stratification, without the use of a ``mixing-length''
parameter, by solving the system of Reynolds momentum equations using the
closure formalism of \citet{Canuto+Dubov:1,Canuto+Dubov:2}. A fixed value of
velocity anisotropy is assumed, and the local convection model is assumed to be
valid for the convectively unstable layer. In accordance with seismic
constraints, overshoot (defined as the amount by which the convectively mixed
zone extends beyond its boundary in local theory) is found to be as low as
about 6 percent of the pressure scale height, and it is not bounded by a
discontinuity from below. 

}

   \maketitle

\section{Introduction}

The layers situated below the convective zone of the Sun are currently
considered to play an important role  in the dynamo mechanism that maintains
the general solar magnetic field \citep{Petrovay:SOLSPA}. Besides, turbulent
mixing processes taking place in this layer are invoked to explain the lithium
deficit of the solar atmosphere and to explain the remaining small
discrepancies between the seismic and standard solar models \citep{Gough+:Sci}.
This turbulence may result partly from convective overshoot, and partly from
rotational instabilities in the solar tachocline. The relative importance of
these two sources of turbulence is an important question.


The first models for the lower overshoot layer were based on nonlocal
extensions of mixing-length theory
\citep{vanBallegooijen:ushoot,Skaley+Stix,Zahn:ushoot,Umezu}. These
predicted\footnote{In fact, adiabatic stratification is explicitly assumed in
some nonlocal MLT models, while in others it is a direct consequence of the
assumption of a perfect correlation between vertical velocity and temperature
fluctuations.} that the thermal stratification of the overshoot layer is nearly
adiabatic, its extent is a significant fraction of the pressure scale height
$H_P$, and it is bounded by a discontinuity in the superadiabaticity
$\Delta \nabla = \nabla - \nablaad$
 from below. ($\nabla = d \ln T / d \ln P$, and $\nablaad =
(\partial \ln T / \partial \ln P)_S$, as usual, $P$, $T$ and $S$ being pressure,
temperature and entropy.)

Such a discontinuity should, however, lead to a noticeable increase in the
amplitude of the oscillatory signal in  the frequency distribution of the
solar oscillations \citep{ChriDa:Tenerife}. Attempts to detect this have
remained unsuccessful
\citep{Monteiro+:seism.undershoot,ChriDa+Monteiro+,Basu+Antia:tachovar}, which
implies that the non-local mixing-length models are incorrect. Thus, either
the overshoot layer is not bounded by a discontinuity from below, or its full
extent is very small, according to \citet{Basu:ushoot} not more than 5\,\% of
$H_P$. It should be stressed, however, that even in the case of a
discontinuity, this limit is only valid for overshoot {\it in the helioseismic
sense,} i.e. for the amount by which convection extends beyond its boundary
calculated in a local model.

Note also that, as \citet{Canuto99} pointed out, from the point of view of
stellar evolution the kinematical or dynamical extent of the overshoot, i.e.
the extent of the mixed layer, is a more important quantity than its thermal
extent, even if the latter were to be adiabatic. Thus throughout this paper we
will define the lower boundary of the overshoot layer at the point where the
kinetic energy is reduced by two orders magnitude compared to its value at
$\Delta\nabla=0$.

A number of numerical simulations have also been carried out to study the
problem, but they have not been conclusive. Owing to the extremely different
dynamical and thermal timescales in the solar interior, the simulations are
either very far from the real solar interior in parameter space
\citep{Roxburgh+Simmons,Hurlburt+Zahn,Singh+:ushoot,Saikia+:ushoot}, or they
never reach a thermally  relaxed state \citep{Andersen:ushoot,Freytag+:ushoot}.

In these circumstances, the most promising way to  model the
overshooting layer is the Reynolds stress method, based on the well known
Reynolds momentum hierarchy of the hydrodynamical equations for a turbulent
medium. 
Our aim in such a model is to determine the radial distribution of the
turbulent kinetic energy $k$, the mean square relative temperature fluctuation
$q$, the normalized energy flux $J$ and the energy dissipation rate
$\varepsilon$ under the convectively unstable region. Note that a
determination of $\varepsilon$ is equivalent to the determination of the
mixing length $l$ as this parameter is usually introduced by defining $l =
k^{3/2} / \varepsilon$. Thus, in this approach the mixing length is determined
consistently within the model and it is not treated as a free parameter. 

The Reynolds moment equations for $k$, $q$ and $J$ were solved for the lower
overshoot layer by \citet{Unno+Kondo}, \citet{Xiong92}, and
\citet{Xiong+Deng:ushoot}, resulting in significant non-adiabatic overshoot.
However, in those calculations $l$ was treated as a free parameter,
arbitrarily set to be equal to $H_P$. The use of an
equation for $\varepsilon$ to get rid of the free parameter was suggested by
\cite{Canuto93} and it has been applied in a simplified $k$--$\varepsilon$
model by \cite{Petrovay:kepsilon}. A fully consistent Reynolds stress
formalism for stellar convection has been developed in a series of papers by
Canuto and co-workers. In this paper we will employ a formalism basically
identical to that of \citet{Canuto+Dubov:1,Canuto+Dubov:2}. This formalism has
been shown to agree well with numerical simulation results for a generic
convective flow with inefficient convection and unit Prandtl number
\citep{Kupka}.

By solving the momentum equations for the realistic solar
stratification, here we present the first model of
the lower overshoot layer without an assumed mixing length
parameter. 

\section{Equations and closure}

We will use the following notations. Any variable $f$ is split into a mean and
a fluctuating part as $f = \overline{f} + f'$. The velocity ${\bf v}$ has only
a fluctuating part. We assume plane parallel geometry and the depth $z$ is
measured from $\Delta \nabla =0$, so the gravity acceleration $g$ is positive.
Beside $\varrho$ for density, we also introduce the notations $q =
\overline{(T' / T)^2}$, $J = \overline{w (T'/T)}$, $w = v_z$, $k =
\overline{v^2} /2 = \overline{w^2} /f_a$ where $f_a$ is an anisotropy
parameter.  

We make the following assumptions:

\begin{itemize}
\item{ $\varrho ' / \varrho \ll 1$}
\item{The Reynolds number $Re \gg 1$, while the Prandtl number $\nu/\chi\ll 1$}
\item{The extent of the overshoot layer $d \ll H_P$, where $H_P$ is the 
pressure scale height.}
\item{The turbulent flow field is characterized by a mild and fixed value of the
anisotropy $f_a\in{\cal O}(1)$. This assumption is made for simplicity only.}
\end{itemize}
With these assumptions we can use the Boussinesq approximation $\varrho ' /
\varrho = - \delta _P (T' / T)$ where $\delta _P$ is an order of unity factor
($\delta_P = 1$ for full ionization). Following standard practice, we will also
neglect the $\overline{{\bf v} \nabla P'}$ term in the equation for $k$.  

The detailed derivation of the Reynolds momentum equations was given by many
authors \citep{Xiong:2,Canuto93,Grossman:1}. With the assumptions and
notations mentioned above, and after customary dimensional modelling of some 
terms, they read
\begin{equation}
\partial _t k = - \partial _z F_k + \delta _p g   J - \varepsilon \label{eq1}
\end{equation}
\begin{eqnarray}
\partial_t J =- \partial_z F_J - C_{qJ}\delta_p gq &&+ C_{kJ}
  f_a \frac{\Delta\nabla}{H_P} k - \frac J{\tau_{p\theta}} \nonumber \\
  &&+ \frac 12\frac\chi T\partial_z^2 (JT) 
          \label{eq2}
\end{eqnarray}
\begin{equation}
\partial _t q = - \partial _z F_q + 2 \frac{\Delta \nabla}{H_P}   J - 
   \frac q{\tau_p} +\frac 12\frac\chi{T^2} \partial_z^2(qT^2)        \label{eq3}
\end{equation}
\begin{equation}
\partial _t \varepsilon = - \partial _z F_{\varepsilon} + C_{J \varepsilon} \delta _p g \frac{\varepsilon}{k}   J - C_{\varepsilon}   \frac{\varepsilon ^2}{k} \label{eq4}
\end{equation}
where the non-local fluxes are 
\begin{equation}
\begin{array}{ll}
F_k = \overline{w v^2/2}, &  F_q = \overline{w (T' / T)^2} \\
F_J = \overline{w^2 T' /T}, & F_{\varepsilon} = \overline{w \varepsilon _l} 
\label{eq5}
\end{array}
\end{equation}
$\varepsilon _l$ being the local dissipation rate. 
For the coefficients we use the following values:
$C_{k J} = 1.0$, $C_{q J} = 1$, $f_a = 1$,
$C_{\varepsilon} = 1.92$,  $C_{J \varepsilon} = 1.44$, $\delta _P = 1$.


These equations are essentially a special case of the equations of
\citet{Canuto+Dubov:1,Canuto+Dubov:2} for a thin layer, and, for
simplicity, with fixed anisotropy. (Note, however, that our definiton of $J$
and $q$ slightly differs from theirs.) Correspondingly, we supplement Eqs.
(\ref{eq1})--(\ref{eq4}) with the closure suggested by \citet{Canuto+Dubov:1}:
\begin{equation}
\partial _z F_k = - \frac{1}{3} \frac{\partial}{\partial z} \left[ \nu _t 
  \Delta ^{-1} \frac{\partial}{\partial z}(K \Delta) \right],
  \label{eq:clo1}
\end{equation}
\begin{equation}
  \partial _z F_{J'} = - \frac{1}{3} \frac{\partial}{\partial z} \left[ 
  \nu _t \Delta ^{-1} \frac{\partial}{\partial z}(\overline{wT'}\Delta) \right] 
   \label{eq:clo2}
\end{equation}
\begin{equation}
 \partial _z F_{1/2 \overline{{T'}^2}} = - \frac{35}{33} \chi ^{-1} 
  \frac{\partial}{\partial z} \left[ \nu _t ^2 \Delta ^{-2} 
  \frac{\partial}{\partial z} \left( \frac{1}{2} \overline{{T'}^2} 
  \Delta ^2 \right) \right] , \label{eq:clo3}
\end{equation}
\begin{equation}
  \partial _z F_\varepsilon  = - \frac{1}{2} \frac{\partial}{\partial z} 
  \left[ \nu _t (1 + \sigma _t ^{-1}) \frac{\partial \varepsilon}{\partial z} 
  \right] ,         \label{eq:clo4}
\end{equation}
where $\Delta = K^2 \varepsilon ^{-4/3}$, $\nu _t = \frac{2}{25} 
{K^2}/{\varepsilon}$, $\sigma _t =0.72$ and $\chi$ is the radiative 
conductivity. Since $\overline{wT'} = J T$ 
and $\overline{{T'}^2} = q T^2$, we can obtain $\partial _z F_J$ and 
$\partial _z F_q$ multiplying $\partial _z F_{J'}$ by $1/T$ and 
$\partial _z F_{1/2 \overline{T'^2}}$ by $2/{T^2}$.

Finally, the timescales $\tau_{p\theta}$ and $\tau_p$ were calculated using
formulae (34) of \citet{Canuto+Dubov:2}.

\section{Numerical solution}
We are looking for stationary solutions of Eqs. (\ref{eq1})--(\ref{eq4}),
in which case the system reduces to an eighth-order system of
ordinary differential equations. The condition of flux equilibrium
\begin{equation}
F_c+F_k+F_r=F_\odot=\,\mbox{const.}   \label{eq:fluxcond}
\end{equation}
where $F_\odot$ is the total solar flux, $F_c$ is the convective heat flux,
$F_k$ is the kinetic energy flux, and $F_r$ is the radiative flux, must also be
coupled to the problem.

The upper boundary conditions are set in a point slightly within the
convectively unstable layer, where we fit the solution to the solution given
by a {\it local\/} convection theory. The equations of local theory are
identical to Eqs. (\ref{eq1})--(\ref{eq4}), except that the non-local
fluxes are not present and the radiative terms are negligible. So these
equations reduce to an algebraic system for our variables. As the system is
redundant, in addition we also use the condition (\ref{eq:fluxcond}) to set
the boundary values. The only free parameter is $\Delta \nabla$ which was
given a small positive value ($\sim 10^{-8}$).

A relaxation method was used to solve this nonlinear, coupled system of
differential equations. 
We start the integration from an arbitrary initial distribution where the
dependent variables are gaussian functions of $z$, of a width comparable to
$H_P$. During integration the following restrictions were imposed:
\begin{itemize}   
\item{Below the point where $k = 0$ or $\varepsilon=0$, all dependent variables
are set to zero.}
\item{Wherever division by $k$ or $\varepsilon$ occurs in the equations we have
used the condition $k \ge 10^{-100}$ and $\varepsilon \ge 10^{-100}$, to avoid
floating point overflow.}
\item{At points where the velocity--temperature correlation  $\hat J\equiv
J/(f_akq)^{1/2} > 1$ or $< -1$, this quantity is set to $1$ or $-1$, 
respectively. Note that this only affects the (physically irrelevant)
intermediate stages during relaxation, while in the final converged solution, 
the unphysical situation mentioned above does not occur.} 
\end{itemize}        

In the initial state we use a linear profile for $\Delta \nabla$  ($\Delta
\nabla = -0.1\,z/H_p$); this crudely represents the actual run of the thermal
stratification below the convectively unstable layer. The iteration goes by
making one ``timestep'' using the full, time-dependent set of Eqs.
(\ref{eq1})--(\ref{eq4}), followed by an instant adaptation of the temperature
profile to ensure the condition of the constancy of the solar flux. This latter
is achieved by calculating the kinetic flux ($F_k = 1/2 \varrho 
\overline{w^2}^{3/2}$) and the convective flux ($F_c = C_p\varrho 
\overline{w T'}$), and subtracting them from the known solar flux to get the
radiative flux
\[ F_r = \frac{4 \sigma}{3 \varrho \kappa} \frac{d T^4}{dr},
\]
from which we can calculate a new temperature stratification ($\kappa$ is the 
opacity and $\sigma$ is the Stefan--Boltzmann constant).

The whole process is repeated until it converges to a stationary state. The
relaxation does not describe a real temporal evolution, as the instant
adaptation of the temperature profile is unphysical, but in this manner we can
circumvent the problem of forbiddingly large radiative/dynamical timescale 
ratios. Thus, the technique is essentially an artificially
accelerated thermal relaxation.

As the layer is thin, the perturbation of the hydrostatic equilibrium is
expected to remain small. The background values of $g(z)$, $\rho(z)$,
$\kappa(z)$ and $H_P(z)$ were thus assumed to be unaffected by overshooting,
and they were taken from a more recent version of the solar model of
\citet{Guenther}.


\begin{figure}[t]
\hspace{-1.1cm}
\includegraphics[width=1.05\linewidth]{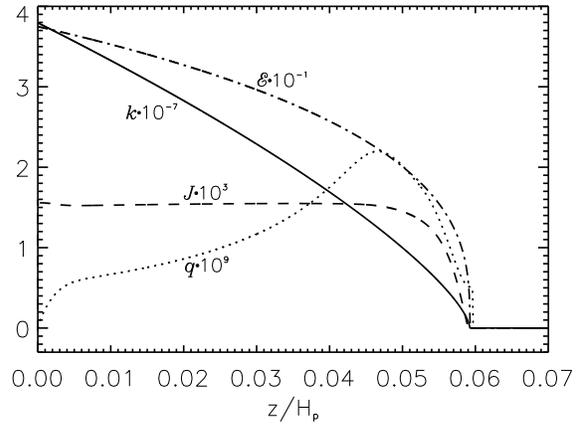}
\caption{Distribution of turbulent kinetic energy $k$, relative squared
temperature fluctuation $q$, convective flux equivalent speed $J$, and  energy
dissipation rate $\varepsilon$ in our model, as functions of depth below the
convectively unstable layer. (All dimensional variables here and in other
figures are given in CGS units.)
\label{fig:1} }
\end{figure}

\begin{figure}[h!]
\hspace{-1.1cm}
\includegraphics[width=1.05\linewidth]{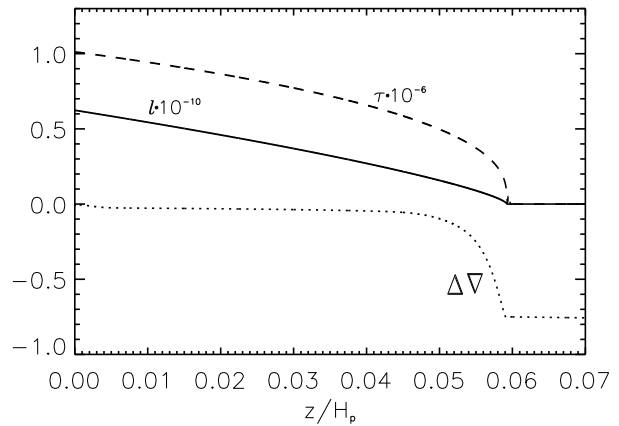}
\caption{Distribution of the turbulent length scale (``mixing length'') $l$,
time scale $\tau$, and superadiabaticity $\Delta\nabla$ in our model, as 
functions of depth below the convectively unstable layer 
\label{fig:2}}
\end{figure}

\section{Results and discussion}

Our stationary solution is presented in
Figs.~1 and 2. The turbulent length and time scales in Fig.~2 are defined
as $l=k^{3/2}/\varepsilon$ and $\tau=k/\varepsilon$.
The thickness of the overshoot layer is strikingly small, below six percent of
$H_P$, and the $\Delta\nabla$ curve does not show a
particularly sharp discontinuity. This is compatible with the available helioseismic evidence \citep{Basu+Antia:tachovar}. It must be stressed that, by design,
our models can only yield the extent of overshoot {\it in the helioseismic
sense} (i.e. the amount by which the convective zone extends beyond its
boundary in local theory). It is quite possible that a proper nonlocal treatment
of the full convective zone would lead to slightly subadiabatic stratification
and to a downward directed $F_c$ (the more conventional criteria for overshoot)
already well above the boundary of the convective zone as defined in local
theory. In this sense, the total extent of the overshoot layer may far exceed
our estimate of the {\it seismic} overshoot.

It is interesting to note that the ``mixing length'' $l$ is found to decrease
continuously towards the bottom of the overshoot layer. This underlines the
incorrectness of the assumption $l=\,$const. in those models that use a free
length parameter.

A doubt may arise regarding the validity of these findings, given that our
model only extends to the overshoot layer itself, assuming the validity of the
local description for the unstable layers above. One sign of the artificiality
of this sudden introduction of nonlocality is the sharp jump in the $J$ curve
from negative to positive values right after $z=0$ (not visible in the figure).
This ``patching'' of our nonlocal overshoot model to the model of the SCZ
computed with a local convection theory also implies that e.g. the full extent
of the convective zone cannot be determined independently from our model.
Indeed, placing the top boundary of the computational domain to a higher level
($z=-0.2 H_P$) we found that overshoot in this case started right at the upper
boundary again. The total extent of this ``seismic'' overshoot, however, showed
only moderate sensitivity to the placement of the upper boundary (resulting in an
overshoot distance of $0.06 H_P$ in the case quoted above). A model
incorporating the full convective zone, with a nonlocal treatment throughout,
is thus clearly a logical future extension of the present model. However, the
solution presented Fig.~3 may give us a reassurance that the small overshoot
is not an artefact of our method. In this plot the stationary  solution of
Eqs. (\ref{eq1})--(\ref{eq4}) is shown, obtained using the simple
down-gradient closure
\begin{eqnarray}
F_k = - \left( k^2/\varepsilon\ \partial_z K  \right)
 &&\quad
F_J = - \left( k^2/\varepsilon\ \partial_z J  \right)
  \\
F_q = - \left( k^2/\varepsilon\ \partial_zq  \right)
 &&\quad
F_\varepsilon = - \left( k^2/\varepsilon\ \partial_z
  \varepsilon  \right)
  \label{eq:dgclo}
\end{eqnarray}
instead of Eqs. (\ref{eq:clo1})--(\ref{eq:clo4}). Note that this
closure is known to be wrong, and it is only used here for comparison. It
is evident that a deep overshoot can also be achieved by our method of
solution; thus, the nature of the closure is the chief responsible for the
smallness of the overshoot.

\begin{figure}[t]
\hspace{-1.1cm}
\includegraphics[width=1.05\linewidth]{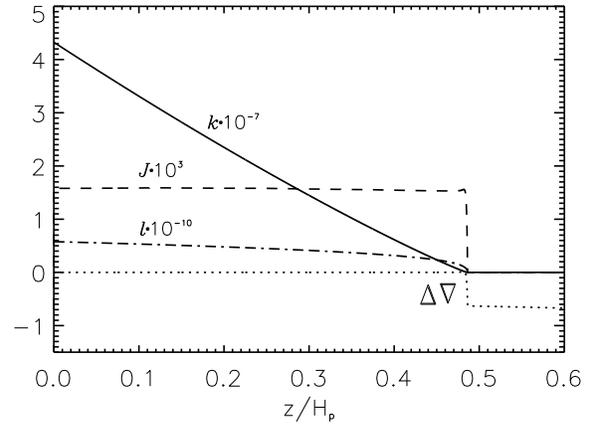}
\caption{Distribution of turbulent kinetic energy $k$, convective flux 
equivalent speed $J$, turbulent length scale $l$, and superadiabaticity
$\Delta\nabla$ in an (unrealistic) solution computed with a simple downgradient
closure, for comparison.
\label{fig:3} }
\end{figure}



Yet we do not claim that our model is the final word concerning overshoot in
the Sun, as several simplifiying assumptions were made both in the model and
in its numerical implementation.  In fact, a definitely better closure for the
non-local fluxes is already available \citep{Canuto:JAS}, and it has been
applied to A stars successfully \citep{Kupka+Montgomery:Astars}. Further work
is also needed to include a consistent treatment of the anisotropy
\citep[cf.][]{Petrovay:GAFD}, to clarify the dependence of the results on
parameter values and closure assumptions, and to extend the domain of
calculation to the unstable layer. 

\acknowledgements
We are grateful to Vittorio Canuto for enlightening discussions and advice.
This work was funded by the OTKA under grants no.~T032462 and T034998.


\begin{thebibliography}{31}
\expandafter\ifx\csname natexlab\endcsname\relax\def\natexlab#1{#1}\fi

\bibitem[{Andersen(1994)}]{Andersen:ushoot}
Andersen, B.~N. 1994, \solphys, 152, 241

\bibitem[{{Basu}(1997)}]{Basu:ushoot}
{Basu}, S. 1997, \mnras, 288, 572

\bibitem[{Basu \& Antia(2001)}]{Basu+Antia:tachovar}
Basu, S., \& Antia, H.~M. 2001, \mnras, 324, 498

\bibitem[{{Canuto}(1993)}]{Canuto93}
{Canuto}, V.~M. 1993, \apj, 416, 331

\bibitem[{{Canuto}(1999)}]{Canuto99}
{Canuto}, V.~M. 1999, \apj, 518, L119

\bibitem[{Canuto(2001)}]{Canuto:JAS}
Canuto, V.~M. 2001, J. Atm.~Sci., 58, 1169

\bibitem[{Canuto \& Dubovikov(1997)}]{Canuto+Dubov:1}
Canuto, V.~M., \& Dubovikov, M. 1997, \apj, 484, L161

\bibitem[{Canuto \& Dubovikov(1998)}]{Canuto+Dubov:2}
---. 1998, \apj, 493, 834

\bibitem[{Christensen-Dalsgaard(1996)}]{ChriDa:Tenerife}
Christensen-Dalsgaard, J. 1996, in The Structure of the Sun, ed. T.~R{oca
  Cort\'es} \& F.~S{\'anchez}, Proc. 6th Canary islands Winter School of
  Astrophysics (Cambridge: Cambridge UP), 47--140

\bibitem[{{Christensen-Dalsgaard} {et~al.}(1995){Christensen-Dalsgaard},
  {Monteiro}, \& {Thompson}}]{ChriDa+Monteiro+}
{Christensen-Dalsgaard}, J., {Monteiro}, M.~J.~P.~F.~G., \& {Thompson}, M.~J.
  1995, \mnras, 276, 283

\bibitem[{Freytag {et~al.}(1996)Freytag, Ludwig, \& Steffen}]{Freytag+:ushoot}
Freytag, B., Ludwig, H.-G., \& Steffen, M. 1996, \aap, 313, 497

\bibitem[{{Gough} {et~al.}(1996){Gough}, {Kosovichev}, {Toomre}, {Anderson},
  {Antia}, {Basu}, {Chaboyer}, {Chitre}, {Christensen-Dalsgaard},
  {Dziembowski}, {Eff-Darwich}, {Elliott}, {Giles}, {Goode}, {Guzik}, {Harvey},
  {Hill}, {Leibacher}, {Monteiro}, {Richard}, {Sekii}, {Shibahashi}, {Takata},
  {Thompson}, {Vauclair}, \& {Vorontsov}}]{Gough+:Sci}
{Gough}, D.~O., {Kosovichev}, A.~G., {Toomre}, J., {et~al.} 1996, Science, 272,
  1296

\bibitem[{{Grossman} {et~al.}(1993){Grossman}, {Narayan}, \&
  {Arnett}}]{Grossman:1}
{Grossman}, S.~A., {Narayan}, R., \& {Arnett}, D. 1993, \apj, 407, 284

\bibitem[{Guenther {et~al.}(1992)Guenther, Demarque, Kim, \&
  Pinsonneault}]{Guenther}
Guenther, D.~B., Demarque, P., Kim, Y.-C., \& Pinsonneault, M.~H. 1992, \apj,
  387, 372

\bibitem[{Hurlburt {et~al.}(1994)Hurlburt, Toomre, Massaguer, \&
  Zahn}]{Hurlburt+Zahn}
Hurlburt, N.~E., Toomre, J., Massaguer, J.~M., \& Zahn, J.-P. 1994, \apj, 421,
  245

\bibitem[{Kupka(1999)}]{Kupka}
Kupka, F. 1999, \apj, 526, L45

\bibitem[{Kupka \& Montgomery(2002)}]{Kupka+Montgomery:Astars}
Kupka, F., \& Montgomery, M.~H. 2002, \mnras, 330, L6

\bibitem[{Monteiro {et~al.}(1994)Monteiro, Christensen-Dalsgaard, \&
  Thompson}]{Monteiro+:seism.undershoot}
Monteiro, M. J. P. F.~G., Christensen-Dalsgaard, J., \& Thompson, M.~J. 1994,
  \aap, 283, 247

\bibitem[{Petrovay(1992)}]{Petrovay:GAFD}
Petrovay, K. 1992, Geophys.\ Astrophys.\ Fluid Dyn., 65, 183

\bibitem[{{Petrovay}(1998)}]{Petrovay:kepsilon}
{Petrovay}, K. 1998, in IAU Symp. 185: New Eyes to See Inside the Sun and
  Stars, Vol. 185, 121--122

\bibitem[{Petrovay(2000)}]{Petrovay:SOLSPA}
Petrovay, K. 2000, in The Solar Cycle and Terrestrial Climate (ESA Publ.
  SP-463), 3--14

\bibitem[{Roxburgh \& Simmons(1993)}]{Roxburgh+Simmons}
Roxburgh, L.~W., \& Simmons, J. 1993, \aap, 277, 93

\bibitem[{Saikia {et~al.}(2000)Saikia, Singh, Chan, Roxburgh, \&
  Srivastava}]{Saikia+:ushoot}
Saikia, E., Singh, H.~P., Chan, K.~L., Roxburgh, I.~W., \& Srivastava, M.~P.
  2000, \apj, 529, 402

\bibitem[{Singh {et~al.}(1995)Singh, Roxburgh, \& Chan}]{Singh+:ushoot}
Singh, H.~P., Roxburgh, I.~W., \& Chan, K.~L. 1995, \aap, 295, 703

\bibitem[{{Skaley} \& {Stix}(1991)}]{Skaley+Stix}
{Skaley}, D., \& {Stix}, M. 1991, \aap, 241, 227

\bibitem[{{Umezu}(1992)}]{Umezu}
{Umezu}, M. 1992, \mnras, 258, 107

\bibitem[{Unno \& Kondo(1989)}]{Unno+Kondo}
Unno, W., \& Kondo, M. 1989, \pasj, 41, 197

\bibitem[{{van Ballegooijen}(1982)}]{vanBallegooijen:ushoot}
{van Ballegooijen}, A.~A. 1982, \aap, 113, 99

\bibitem[{{Xiong}(1980)}]{Xiong:2}
{Xiong}, D.~R. 1980, Chinese Astron. Astrophys., 4, 234

\bibitem[{{Xiong} \& {Chen}(1992)}]{Xiong92}
{Xiong}, D.~R., \& {Chen}, Q.~L. 1992, \aap, 254, 362

\bibitem[{{Xiong} \& {Deng}(2001)}]{Xiong+Deng:ushoot}
{Xiong}, D.~R., \& {Deng}, L. 2001, \mnras, 327, 1137

\bibitem[{{Zahn}(1991)}]{Zahn:ushoot}
{Zahn}, J.-P. 1991, \aap, 252, 179

\end{thebibliography}

\end{document}